\documentclass[]{aa}

\usepackage{graphicx}
\usepackage{txfonts}
\usepackage{hyperref}
\usepackage{graphicx}
\usepackage{gensymb}
\usepackage{multirow}
\usepackage{amsmath}
\usepackage{tablefootnote}
\usepackage{placeins}

\begin{document}

   \title{A giant thin stellar stream in the Coma Galaxy Cluster}

   \authorrunning{Román et al.}

   \author{Javier Román\inst{1,2,3}\thanks{   \email{jromanastro@gmail.com}}
   \and
   R.~Michael~Rich \inst{4}
   \and
   Niusha Ahvazi \inst{5,6}
   \and
   Laura V. Sales \inst{5}
   \and
   Chester Li \inst{4,7}
   \and
   Giulia Golini \inst{2,3}
   \and
   Ignacio Trujillo \inst{2,3}
   \and
   Johan~H.~Knapen \inst{2,3}
   \and 
   Reynier~F.~Peletier \inst{1}
   \and
   Pablo~M.~Sánchez-Alarcón \inst{2,3}
   }

   \institute{Kapteyn Astronomical Institute, University of Groningen, PO Box 800, 9700 AV Groningen, The Netherlands
   \and Instituto de Astrof\'{\i}sica de Canarias, c/ V\'{\i}a L\'actea s/n, E-38205, La Laguna, Tenerife, Spain
   \and Departamento de Astrof\'{\i}sica, Universidad de La Laguna, E-38206, La Laguna, Tenerife, Spain
   \and Department of Physics \& Astronomy, University of California Los Angeles, 430 Portola Plaza, Los Angeles, CA 90095-1547, USA
   \and Department of Physics and Astronomy, University of California, Riverside, 900 University Avenue, Riverside, CA 92521, USA
   \and Carnegie Observatories, 813 Santa Barbara Street, Pasadena, CA 91101, USA
   \and Department of Astronomy, University of Washington, 3910 15th Avenue, NE,  Seattle, WA, 98195, USA
   }
   \date{\today}

 
  \abstract
   {The study of dynamically cold stellar streams reveals information about the gravitational potential where they reside and provides important constraints on dark matter properties. However, their intrinsic faintness makes detection beyond Local environments highly challenging. Here we report the detection of an extremely faint stellar stream ($\mu_{g,max}$~$=$~29.5~mag~arcsec$^{-2}$) with an extraordinarily coherent and thin morphology in the Coma Galaxy Cluster. This {Giant~Coma~Stream} spans $\sim$510~kpc in length and appears as a free-floating structure located at a projected distance of 0.8 Mpc from the centre of Coma. We do not identify any potential galaxy remnant or core, and the stream structure appears featureless in our data. We interpret the {Giant~Coma~Stream} as being a recently accreted, tidally disrupting {passive} dwarf. {Using the Illustris-TNG50 simulation, we identify a case with similar characteristics, showing that, although rare, these types of streams are {predicted} to exist in $\Lambda$-CDM. Our work unveils the presence of free-floating, extremely faint and thin stellar streams in galaxy clusters}, widening the environmental context for their promising future applications in the study of dark matter properties.}

   \keywords{Galaxies: clusters: Coma, Galaxies: evolution, Galaxies: interactions, Galaxies: photometry}

   \maketitle
%

\section{Introduction}

Extremely low surface brightness features in the form of stellar streams and haloes provide crucial information for understanding the hierarchical model of galaxy evolution in a $\Lambda$-CDM framework \citep[][]{2005ApJ...635..931B, 2008ApJ...689..936J, 2010MNRAS.406..744C, 2014MNRAS.444..237P, 2019MNRAS.485.2589M, 2022MNRAS.510.4208R, 2023MNRAS.520.3767G}. These structures are created from the continuous accretion of satellites, and representative examples are found in the environments of the Local Group \citep[e.g., ][]{2006ApJ...642L.137B, 2018ApJ...868...55M, 2021ApJ...914..123I, 2023A&A...670L...2D} and of nearby galaxies \citep[e.g., ][]{2010ApJ...714L..12M, 2016ApJ...823...19C} by counting individual stars.

{The availability of high-precision stellar positions and velocities,} mainly in the Milky Way environment, has promoted intense research in this field \citep[see a review by ][]{2020ARA&A..58..205H}. One of the applications of the study of stellar streams, given their extremely low stellar density, is their availability to reveal the gravitational potential in which they reside \citep[e.g., ][]{1999ApJ...512L.109J, 1999ApJ...526..607D, 2013MNRAS.433.1826S, 2021MNRAS.502.4170R, 2022ApJ...941...19P}, especially for the case of cold stellar streams \citep{2018ApJ...867..101B}. {Because of their} coherent stellar motion, any nearby perturbation by a conspicuous low-mass potential would be easily identifiable in parameter space \citep[e.g., ][]{2002MNRAS.332..915I, 2012ApJ...748...20C, 2015MNRAS.454.3542E}. This is particularly interesting for the case of a potential perturbation by dark matter subhaloes, which could provide relevant information about the properties of dark matter {in already known streams in the Milky Way such as Pal 5 or GD-1} \citep[][]{2016MNRAS.463..102E, 2019MNRAS.484.2009B, 2020ApJ...891..161I, 2021JCAP...10..043B}.

Because of the large amount of information contained in stellar streams, there have been sustained efforts to explore these structures in external galaxies to obtain a larger environmental and statistical context \citep[][]{2009AJ....138.1417T, 2010AJ....140..962M, 2012Natur.482..192R, 2015MNRAS.446..120D, 2021A&A...654A..40T, 2023A&A...671A.141M}. However, {resolving individual stars is feasible out to only a few Mpc, and the difficult observational challenges of large-area, low surface brightness photometry \citep[][]{2017ASSL..434..255K, 2019arXiv190909456M} mean that the very low surface brigthness regimes at which cold stellar streams have been identified in the Milky Way are basically unexplored at greater distances, and most of the expected faint stellar streams remain undetected \citep[][]{2022MNRAS.513.1459M}.}


With the imminent arrival of the new generation of instrumentation and deep optical surveys such as Euclid \citep{2022A&A...662A.112E}, the Rubin Observatory \citep{2019ApJ...873..111I} or the Nancy Grace Roman Space Telescope \citep{2019arXiv190205569A}, along with significant technical efforts \citep{2022A&A...657A..92E, 2023MNRAS.519.4735S, 2023MNRAS.520.2484K}, it is expected that the surface brightness limits to be reached will increase by orders of magnitude the numbers of stellar streams detected in external galaxies \citep{2019ApJ...883...87P}. This will provide a broader environmental context and a greater understanding of the processes involved, yielding invaluable information in both galactic evolution models and near-field cosmology.

We present the first results of an extensive observational campaign to explore the Coma cluster at the ultra-low surface brightness regime: The HERON Coma Cluster Project. {The Coma cluster is one of the most-studied extragalactic objects, and of particular historical significance \citep[see a historical review by ][]{1998ucb..proc....1B}, being the site, for example, of the first evidence of dark matter \citep{1933AcHPh...6..110Z} or intracluster-light \citep{1951PASP...63...61Z}. The existence of galactic debris in Coma \citep{1998MNRAS.293...53T, 1998Natur.396..549G} and other clusters \citep{1999AJ....117...75C, 2000MNRAS.314..324C, 2017ApJ...834...16M} is direct evidence of the processes of galactic cannibalism and accretion giving rise to the building of the intragroup and intracluster light haloes \citep{2009ApJ...699.1518R, 2015MNRAS.448.1162D, 2022NatAs...6..308M}. As Coma is one of the most massive local clusters with intense merger activity \citep[e.g., ][]{2019A&A...622A.183J, 2020ApJ...894...32G}, it is an {ideal setting} to carry out an exploration of this type of structures, providing crucial information on the environmental and interaction processes undergone in clusters.}

{Here, we report the discovery of the Giant Coma Stream, a 510 kpc long, extremely faint, thin, and free-floating stellar stream in the outskirts of the Coma galaxy cluster.  It is several times longer than the two previously-identified stellar streams in the Coma cluster \citep{1998MNRAS.293...53T, 1998Natur.396..549G} or other clusters \citep{1999AJ....117...75C, 2000MNRAS.314..324C, 2017ApJ...834...16M}, while being orders of magnitude lower surface brightness and total mass.  We show that this new stream is consistent with features found in numerical simulations of hierarchical cluster formation (Illustris TNG).} We assume a distance of 100 Mpc for Coma \citep{2001ApJ...557L..31L}, corresponding with a distance modulus of m-M~=~35.0~mag and spatial scale of 0.462~kpc~arcsec$^{-1}$ {using cosmological parameters from \cite{2020A&A...641A...6P}}. We use the AB photometric system throughout this work.



\section{Observations and detection}

\subsection{HERON data}\label{sec:HERON}

We carried out an extensive observational campaign of deep observations in the Coma Cluster in $g$ and \textit{r} bands with the 0.7m Jeanne Rich telescope. This telescope is mainly dedicated to the Halos and Environments of Nearby Galaxies (HERON) Survey \citep[][]{2019MNRAS.490.1539R}, and is designed to be particularly efficient in the low surface brightness regime. {It has a single Fingerl Lakes Instruments ML09000 detector with $3048^2$ 12$\mu \rm m$ pixels with scale 1.114 arcsec/pix, allowing a $57\times 57$ arcmin field of view. Further technical description of the instrumentation is detailed by \cite{2017IAUS..321..186R}. Observations in this work} are part of the HERON Coma Cluster Project (Román et al., in prep.). The goal is to carry out a study of the diffuse light of the Coma cluster to unprecedented limits in surface brightness.

Observations were conducted in the spring and summer of 2019 for the \textit{g} band and 2020 for the \textit{r} band. A total of 461 and 579 exposures of 300 sec were taken in the \textit{g} and \textit{r} bands respectively. Observational conditions were mostly dark. Dithering steps of tens of arcmin were performed with the aim of covering an area of 1.5$\times$1.5 degrees centred on Coma. The data reduction was performed by standard subtraction of combined superbias and superdarks for each night from the science images. {Flat-fields} were constructed by combining and normalizing the heavily masked bias-subtracted and dark-subtracted science images. Astrometry was performed on all images individually with the \texttt{Astrometry.net} software package \citep{2010AJ....139.1782L} and SCAMP \citep{2006ASPC..351..112B}. The images were then reprojected onto a common astrometrical grid with SWARP \citep{2002ASPC..281..228B} and photometrically calibrated with the Dark Energy Camera Legacy Survey \citep[DECaLS;][]{2019AJ....157..168D} as a reference. {The images were combined with a very conservative sky fitting based on Zernike polynomials {to preserve} the information at low surface brightness. We used orders between 1 and 4 depending on the degree of gradients in the images, and always using the lowest possible order that is able to fit the gradients of each individual image.}

The total exposure time was 38.4h and 48.25h in the \textit{g} and \textit{r} bands, respectively. The maximum surface brightness limits in these data are 30.1 and 29.8 mag arcsec$^{-2}$ [3$\sigma$, 10$\times$10"] in \textit{g} and \textit{r} bands \citep[see definition by][]{2020A&A...644A..42R}. In the region of interest here, the surface brightness limits reach 29.5 mag arcsec$^{-2}$ [3$\sigma$, 10$\times$10"] in both \textit{g} and \textit{r} bands. The seeing conditions were not restrictive as we were interested in the diffuse light, producing a final seeing of approximately 3 arcsec in both \textit{g} and \textit{r} bands. Further details about this Project and data will be published soon. 

\begin{figure*}
\centering
        \includegraphics[width=0.98\textwidth]{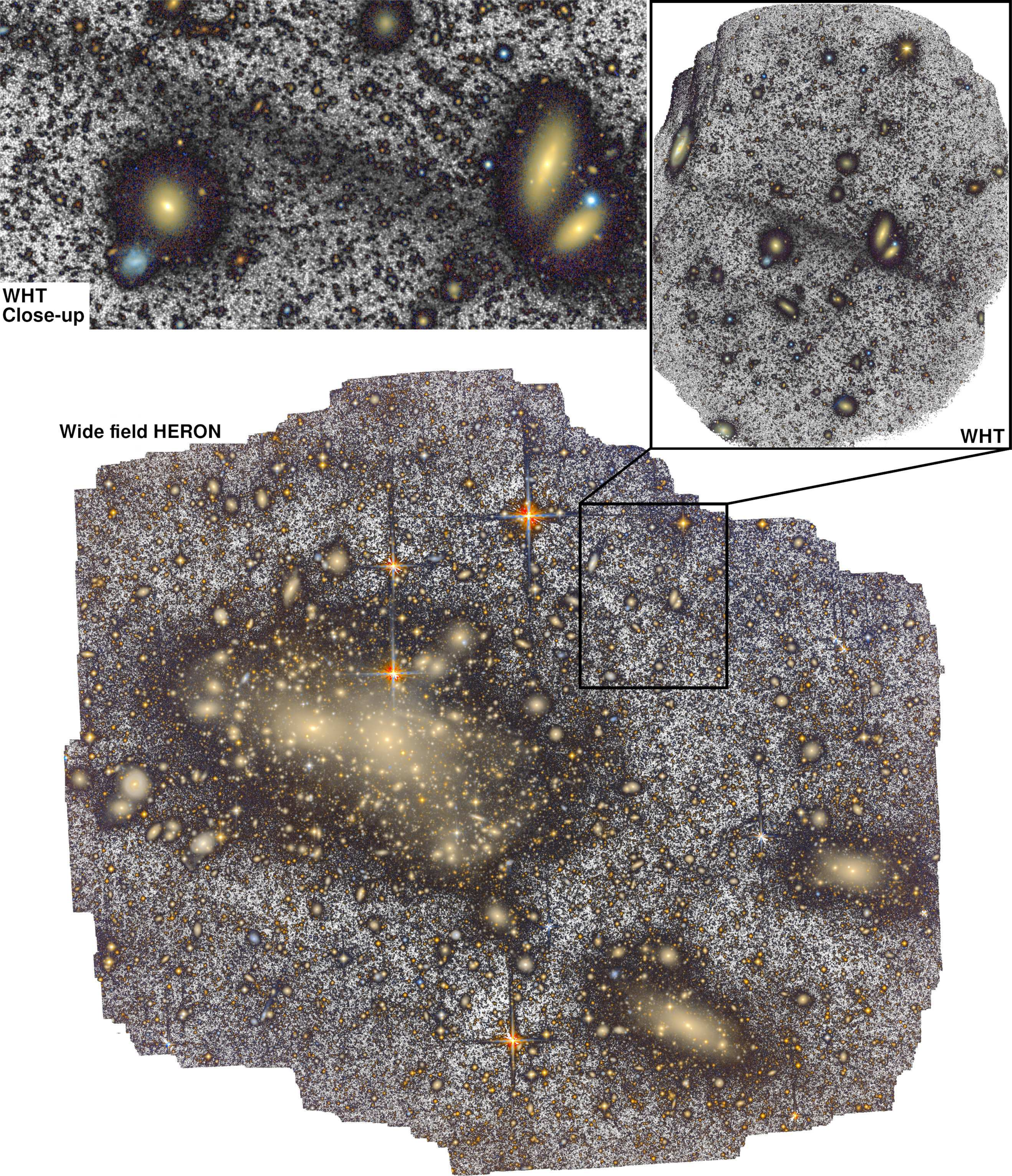}
    \caption{a) Colour composite image from the HERON Coma Cluster Project in the \textit{g} and \textit{r} bands. The dark background is constructed with the sum image. b) William Herschel telescope image of the {Giant~Coma~Stream} in {the $L$} luminance (dark background) on a colour image using \textit{g} and \textit{r} bands from the DECaLS survey. c) Zoom-in on the central region of the image in b).}
    \label{fig:Panel}
\end{figure*}

A preliminary {visual} analysis of these data allowed us to identify an extremely faint feature with very thin morphology (see~Fig.~\ref{fig:Panel},~a). Its detected length and width are approximately 18.5 and 1 arcmin respectively.

\subsection{WHT data}

To get confirmation of this feature, we carried out follow-up observations with the 4.2m William Herschel Telescope. We used the PF-QHY Camera with a field of view of 7.1'$\times$10.7' and pixel scale of 0.2667 arcsec/pixel. We used a $L$ luminance filter which is basically a UV-IR blocker with a high efficiency of 95\% in the range 370-720 nm, thus covering the \textit{g} and \textit{r} SDSS filters. {The purpose of using a wide luminance band is to achieve maximum detection power given the extremely low surface brightness of this feature.} A total of 200 individual 180-second exposures {was} obtained on June 4, 6, 11, and 14, 2021 under seeing conditions of approximately 1.2 arcsec. {The data processing was similar to the one used for the HERON data.} Throughout the observations we carried out extensive dithering, {with maximum separation of about 5 armin (half of the field of view),} in order to obtain high efficiency in the low surface brightness regime, minimizing the presence of gradients and allowing us to build a flat-field with the science images.

Due to the {extensive} dithering of the observations, the exposure time varies {with position}. In Fig. \ref{fig:Mapas_depth} {in the Appendix,} we show the exposure time and equivalent depth along the footprint of the observations. In the maximum exposure region of 10 hours the limiting surface brightness reaches 31.4 mag arcsec$^{-2}$ [3$\sigma$, 10$\times$10"] in the $L$ luminance band. A relatively large portion of about 6x8 arcmin of the area of the image, including the central region of the stream, has a surface brightness limit of 31.0 mag arcsec$^{-2}$ [3$\sigma$, 10$\times$10"]. The resolution is 1.2 arcsec of full width half maximum.

{The WHT luminance band image confirms the existence of the narrow feature identified in the HERON data, achieving deeper surface brightness limits and higher resolution} (see Fig.~\ref{fig:Panel},~b~and~c) over the central region (14 $\times$ 17 arcmin) of the stream. We searched {visually} for potential remnants or overdensity from a parent galaxy. {However, the {stream} appears completely featureless, especially in the higher resolution and deeper WHT image that samples the central region.}

\section{Association with the Coma Cluster}

The detection of this stream in three images with two different telescopes allows us to confirm its discovery and to rule out a possible residual flux due to instrumentation or data processing. 

The presence of globular clusters usually traces the trajectories of streams \citep[][]{2019Natur.574...69M}, which could also provide an estimate of their distance by the peak of the globular cluster luminosity function \citep[][]{2012Ap&SS.341..195R}. However, the HERON data do not have enough point-source depth, and do not allow colour filtering with only the $g$ and $r$ bands. There are no HST data available over the area, and the only multi-band data available from DECaLS have {insufficient point source depth, making globular cluster detection infeasible until better data exist.} {In this section we perform the analysis to ensure that the detected feature is indeed a stellar stream located in Coma.}

\subsection{Potential resolved stars}

{We explore the possibility of the existence of an overdensity of point-like sources over the stream region, which would indicate a close distance}. For this we use \texttt{SExtractor} \citep[][]{2006ASPC..351..112B} with a detection threshold of 1 sigma. We do not apply any magnitude criteria, other than selecting sources compatible with point-like sources, and for this we select sources with a stellarity index higher than 0.5. 

In Fig. \ref{fig:Opt_IR}, middle panel, we plot the density of point-like sources that we can associate with low-luminosity stars. Interestingly, this stream is not resolved into stars in the WHT data, appearing as an extremely smooth and low surface brightness feature. Considering that this feature is not resolved {into} stars, we follow the analysis by \cite{2012MNRAS.421..190Z} to impose a lower limit on distance. Taking into account our best resolution data with a seeing of 1.2 arcseconds and a surface brightness of approximately 29 mag arcsec$^{-2}$ (to be detailed in later sections), this implies that the feature must be at a distance of at least 1 Mpc to appear unresolved {into stars, therefore not associated with any structure of the Local Group.}

\subsection{Far-IR counterpart}\label{sec:IR}

In order to explore the possibility that the stream may be a trace of dust from the interstellar medium, Galactic cirrus of the {Milky Way}, we used data available from the Herschel space telescope at 250 $\mu$m \citep[][]{2010A&A...518L...1P}. Due to its low temperature, Galactic cirrus {is} efficiently detected in far-IR or submillimeter bands \citep[][]{1984ApJ...278L..19L, 2010ApJ...713..959V}. The Herschel data at 250 $\mu$m offer a significant advantage over other data such as IRAS \citep[][]{1984ApJ...278L...1N} or the 857 Ghz band of the Planck observatory \citep[][]{2011A&A...536A..24P} both in detection power, but also, importantly for our case, in resolution.

\begin{figure}
\centering
        \includegraphics[width=1.0\columnwidth]{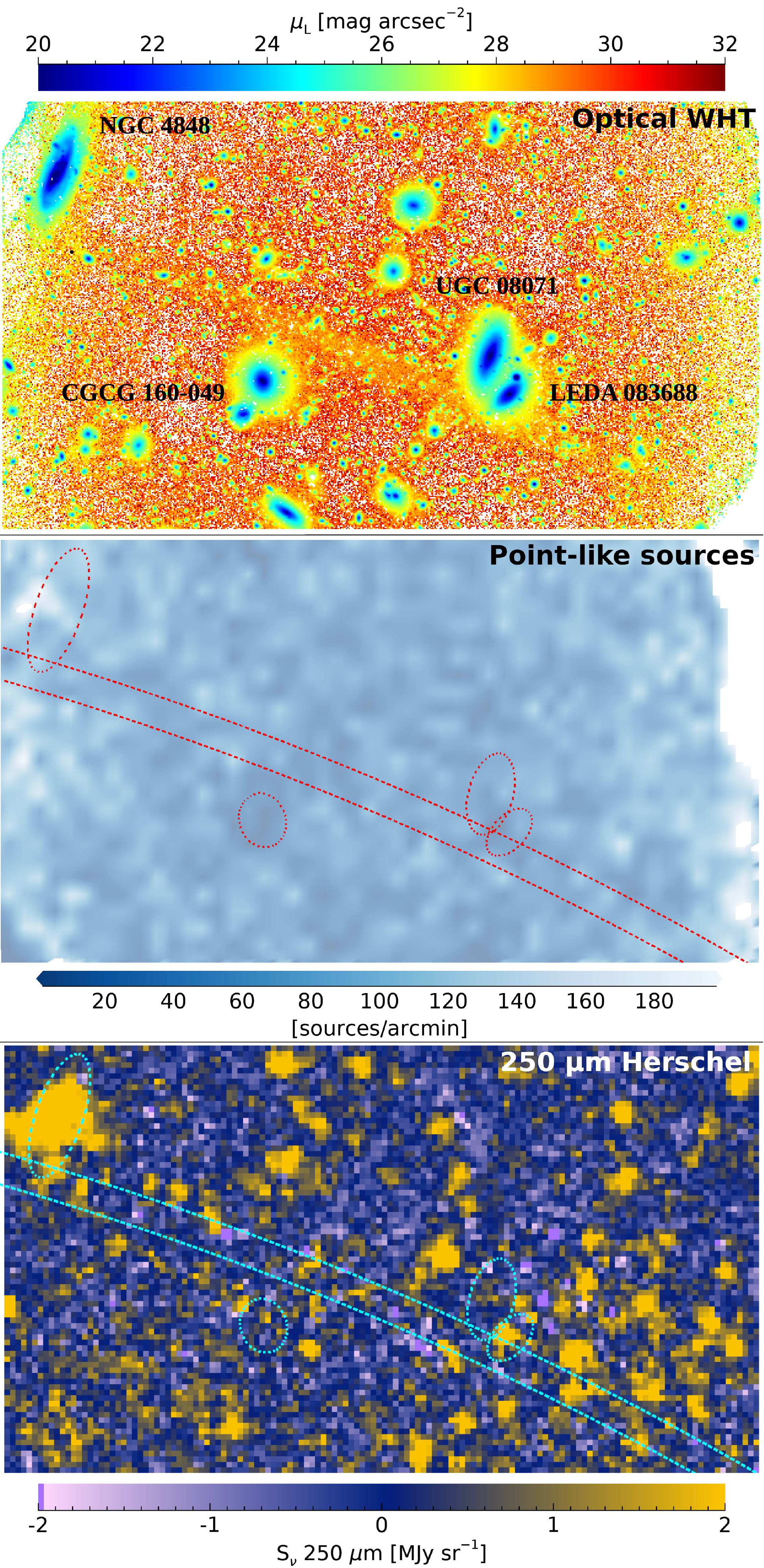}
    \caption{{Comparison between optical, {point-like sources} density and far-IR emission in the central region (13.5$\times$7.6 arcmin). Top panel: WHT Image in the luminance $L$ filter colour coded in surface brightness according to the upper colour bar. Middle panel: Density of detected {point-like sources in the region with WHT data}. Lower panel: 250 $\mu$m counterpart with the Herschel space telescope. The image is colour coded in $S_{\nu}$ according to the lower colour bar. The dashed lines mark elliptical surface brightness contours of approximately $\mu_{L}$ = 26 mag arcsec$^{-2}$ for the galaxies named in the top panel, and $\mu_{L}$ = 29.5 mag arcsec$^{-2}$ for the location of the {Giant~Coma~Stream.}}}
    \label{fig:Opt_IR}
\end{figure}

In Fig. \ref{fig:Opt_IR}, lower panel, we show a comparison between our optical data and the Herschel 250 $\mu$m band of the same field. The images have been stretched {in the optical} to provide maximum contrast for the detection of the faintest sources. {The FWHM of the 250 $\mu$m Herschel data is 17.6 arcsec, smaller than the detected stream width in optical bands of 1 arcmin.} {Visually,} there is no emission trace in the 250 $\mu$m band, {either} in the region where the stream is located {or} adjacent to it. The only detectable emission in 250 $\mu$m is that corresponding to galaxies along the line of sight, among which we can highlight the strong emission from the late-type galaxy NGC 4848. 

{In order to quantify the 250 $\mu$m emission over the region of the stream, we calculated the total flux in the aperture defined by the dashed lines shown in Fig. \ref{fig:Opt_IR} and outside this region. This aperture is defined using HERON optical data, and will be detailed in Sec. \ref{sec:photo}. This is done on the heavily masked image, avoiding integrating regions corresponding to external sources that appear in the image. The average sky background value over the region of the stream is $S_{\nu,250\mu m}$~=~$-$0.02~$\pm$~0.04~MJy sr$^{-1}$ and outside this region is $S_{\nu,250\mu m}$~=~0.02~$\pm$~0.01~MJy sr$^{-1}$. Therefore, the stream has no emission in the 250 $\mu$m band, at least up to the value defined by the sky background noise of these data.}

The absence of a 250 $\mu$m counterpart to the optically detected stream suggests that the Galactic cirrus emission is unlikely as an explanation for the origin of the feature. Additional reasons to rule out a Galactic cirrus contamination are that this celestial region is very close to the Galactic pole ($b$~=~88$^{\circ}$), with a low extinction by dust (A$_r$ = 0.02 mag) \citep{2011ApJ...737..103S} and that the stream is nearly one dimensional, not having the expected fractal morphology of cirrus in optical wavelengths \citep{2021MNRAS.508.5825M}.

\subsection{Environmental analysis}

The extremely low surface brightness of this feature makes it {highly challenging} to obtain spectroscopic information with which to measure its redshift. {We have not found any gas detections in the region using the NASA/IPAC Extragalactic Database, nor counterparts in recent work by \cite[][]{2022ApJ...933..218B}.}

Assuming that the feature is indeed a stellar stream, its elongated morphology suggests that it must be a tidal feature and therefore should be associated with some structure that impacts it gravitationally. Considering its large apparent size, beyond the obvious possible association with the adjacent Coma cluster, we explored a potential association with some nearby structures in close projection. 

\begin{figure*}
\centering
        \includegraphics[width=0.8\textwidth]{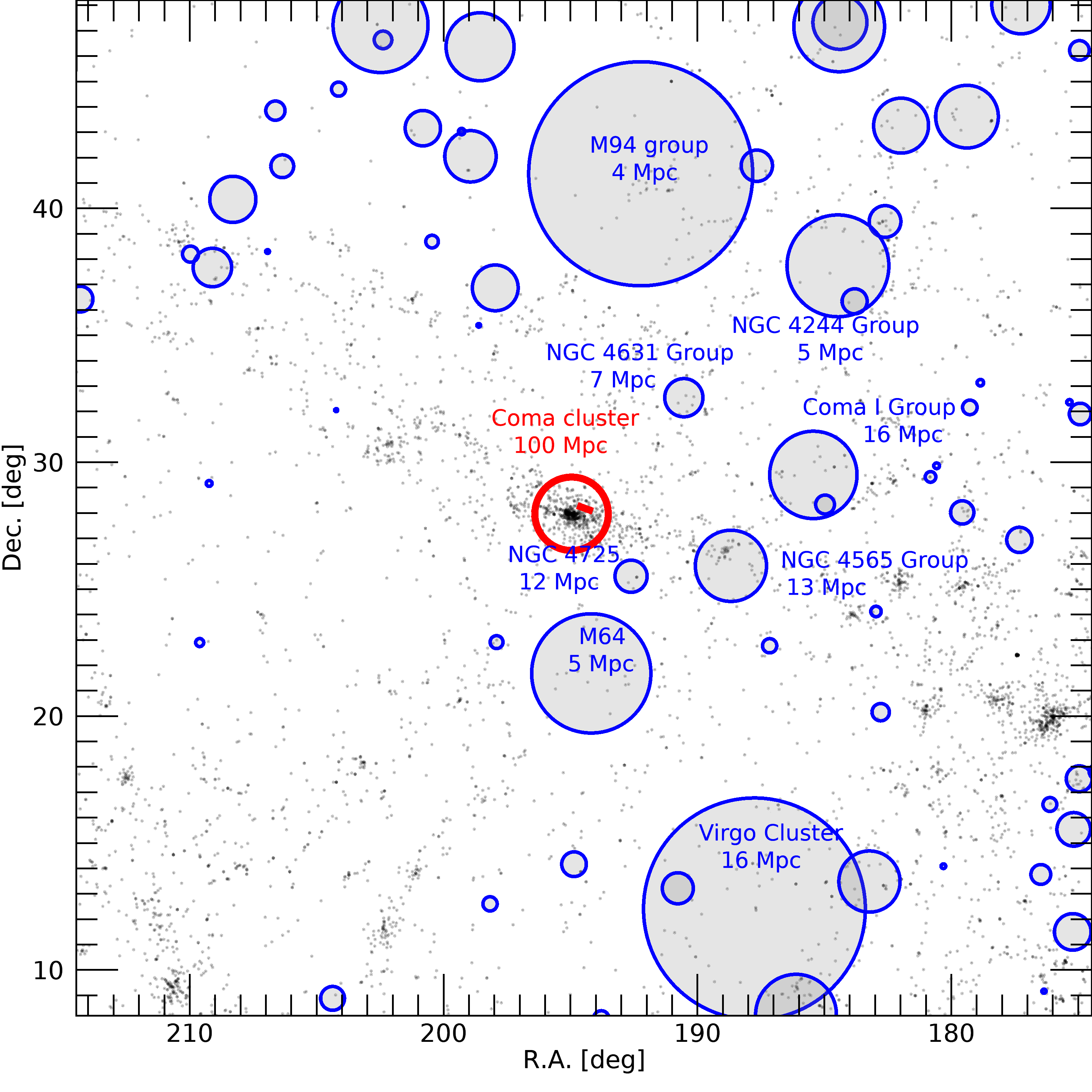}
    \caption{Graphical representation of galactic associations in the line of sight of the {Giant Coma Stream}. Groups and clusters of local galaxies (V$_{rad}$~$<$~3500 km s$^{-1}$) and their apparent virial radii are represented with gray circles. Galaxies with redshift between 3500~km~s$^{-1}$~$<$~V$_{rad}$~$<$~10000~km~s$^{-1}$ are represented with black dots. The apparent Coma virial radius is represented with a red circle. The position and longitude of the {Giant Coma Stream} is represented with a red segment.}
    \label{fig:Environment}
\end{figure*}

In Fig.~\ref{fig:Environment} we show a celestial coordinate map with large-scale structures in the line of sight of the stream. We make use of the catalogue of nearby groups within 3500~km~s$^{-1}$ by \cite{2017ApJ...843...16K} to {identify} these structures. \cite{2017ApJ...843...16K} also offers estimates of the virial radius of these structures. 

We plot all groups together with their virial radii in Fig.~\ref{fig:Environment}. We convert from physical to projected coordinates using the same cosmological parameters as those in \cite{2017ApJ...843...16K} with H$_{0}$ = 75 km s$^{-1}$ Mpc$^{-1}$. Where a distance is not available for a group we use a direct conversion by Hubble's law using the radial velocity. We find that only sparse groups do not have measured distances, and in any case, they are not close in projection to the Coma Cluster. As a complement, and given that the \cite{2017ApJ...843...16K} catalogue only {contains} groups up to 3500 km s$^{-1}$, we plot galaxies with radial velocities between 3500 and 10000 km s$^{-1}$ obtained from the NASA/IPAC Extragalactic Database as black points. This upper limit is motivated as the maximum radial velocity of the galaxies contained in the central region of the Coma cluster. To represent the virial radius of the Coma cluster we rely on \cite{2022NatAs.tmp..148H}, giving 2.4 Mpc with H$_{0}$ = 75 km s$^{-1}$ Mpc$^{-1}$. We also plot as a red segment the length and location of the stream.

We can comment on some nearby structures. First, the Virgo Galaxy Cluster, {which} could be a potential host for the stream, is located at approximately 13 degrees {projected
separation from Coma}. With a calculated virial radius of 1.3 Mpc for the Virgo Cluster, this corresponds to an apparent size of 4.4 degrees, the stream would be therefore located at a projected distance of 3.9 times the virial radius of the Virgo Cluster. The M94 group is the other structure with a large apparent size, having a virial radius of 365 kpc, equivalent to 4.4 degrees. The stream is located at 3 times this virial radius of the M94 Group. NGC 4725, located at a distance of 12 Mpc, has a virial radius of 143 kpc equivalent to 0.63 degrees, with the stream at 5.1 times its virial radius. Next, the group of NGC 4565, located at a distance of 13 Mpc, has a virial radius of 343 kpc equivalent to 1.4 degrees, leaving the stream at more than 4.4 times its virial radius. Finally, the closest structure in terms of virial radii is M64, located at a distance of 5 Mpc, with a virial radius of 224 kpc, equivalent to 2.34 degrees. The stream would therefore be projected at a distance of 2.8 times this virial radius.

We explored the possibility that the stream could be caused by a disturbance from a galaxy that overlaps with it in the line of sight. In Fig. \ref{fig:Opt_IR} the most prominent candidates are identified by name. In particular, the pair of overlapping galaxies UGC~08071 and LEDA~083688 could be candidates to produce the stream. However, our very deep images do not show any sign of asymmetry in them. Furthermore, the radial velocities of UGC~08071 and LEDA~083688 are 6933~±~2~km~s$^{-1}$ and 8169~±~2~km~s$^{-1}$ respectively, thus located more than 1200~km~s$^{-1}$ apart in velocity space (the radial velocity of galaxies in Coma ranges from 4000 to 10000 km~s$^{-1}$). This makes it highly unlikely that they could have interacted at such a high velocity while leaving no asymmetry in the time during which the stream has expanded. The other galaxies in the line of sight also show no appreciable asymmetry that would be compatible with the occurrence of the stream. {We note that for a tidal force calculation between galaxies in the line of sight, a distance value between them is necessary. However in this case it is not feasible considering that we are in a clustered environment, and that very close projections between galaxies might only be apparent. This would mean that galaxies in close projection could be separated by tens or even hundreds of kpc. We therefore consider the potential presence of disturbance in the outer parts of galaxy haloes to be the most realistic approach to denote the presence of interaction between galaxies.}

{We further investigate a possible counterpart to a radio relic. Radio relics are diffuse structures detected in radio by synchrotron emission frequently found in interacting galaxy clusters \citep[see a review by ][]{2019SSRv..215...16V}. Notably, these radio relics are often found in the peripheral regions of galaxy clusters, with morphologies similar to those of the stream \citep[][]{2012A&ARv..20...54F, 2015ApJ...802...46J}. Although these radio relics have no optical counterpart in previous works, due to the extreme surface brightnesses that we explore in this work, it is interesting to look for a counterpart to a radio relic. However, we do not identify a counterpart in the deepest and most recent data in Coma by \citep[][]{2022ApJ...933..218B}.}

From this analysis, we conclude that the stream is embedded within the region of influence or virial radius of the Coma Galaxy Cluster, and does not overlap with the virial radius of any other structure in the line of sight. No other galactic association, such as a group or cluster, overlaps with the location of the stream, and there is no galaxy in the line of sight that could be an origin for the stream. This indicates that the stream is most likely associated with Coma and that there is no other particular galaxy or galactic association that can explain its clearly disturbed morphology.

\section{Analysis}


\subsection{Photometry}\label{sec:photo}

To better constrain the nature and possible origin of the {Giant~Coma~Stream}, we perform the photometric analysis with the two different datasets available to us. The HERON image provides two photometric bands \textit{g} and \textit{r} with surface brightness limits of 29.5 mag arcsec$^{-2}$ [3$\sigma$, 10"$\times$10"] in \textit{g} and \textit{r} bands respectively in the region of interest, with a spatial resolution of $\approx$ 3 arcsec FWHM and completely covering the region. The WHT data partially covers the central region of the stream with an average depth of 31.0 mag arcsec$^{-2}$ [3$\sigma$, 10"$\times$10"] in {the luminance band} and a spatial resolution of approximately 1.2 arcsec of FWHM. Given the different characteristics of the two datasets, we perform a different analysis on {each}, focusing on obtaining different photometric properties.

\begin{figure*}
\centering
        \includegraphics[width=0.8\textwidth]{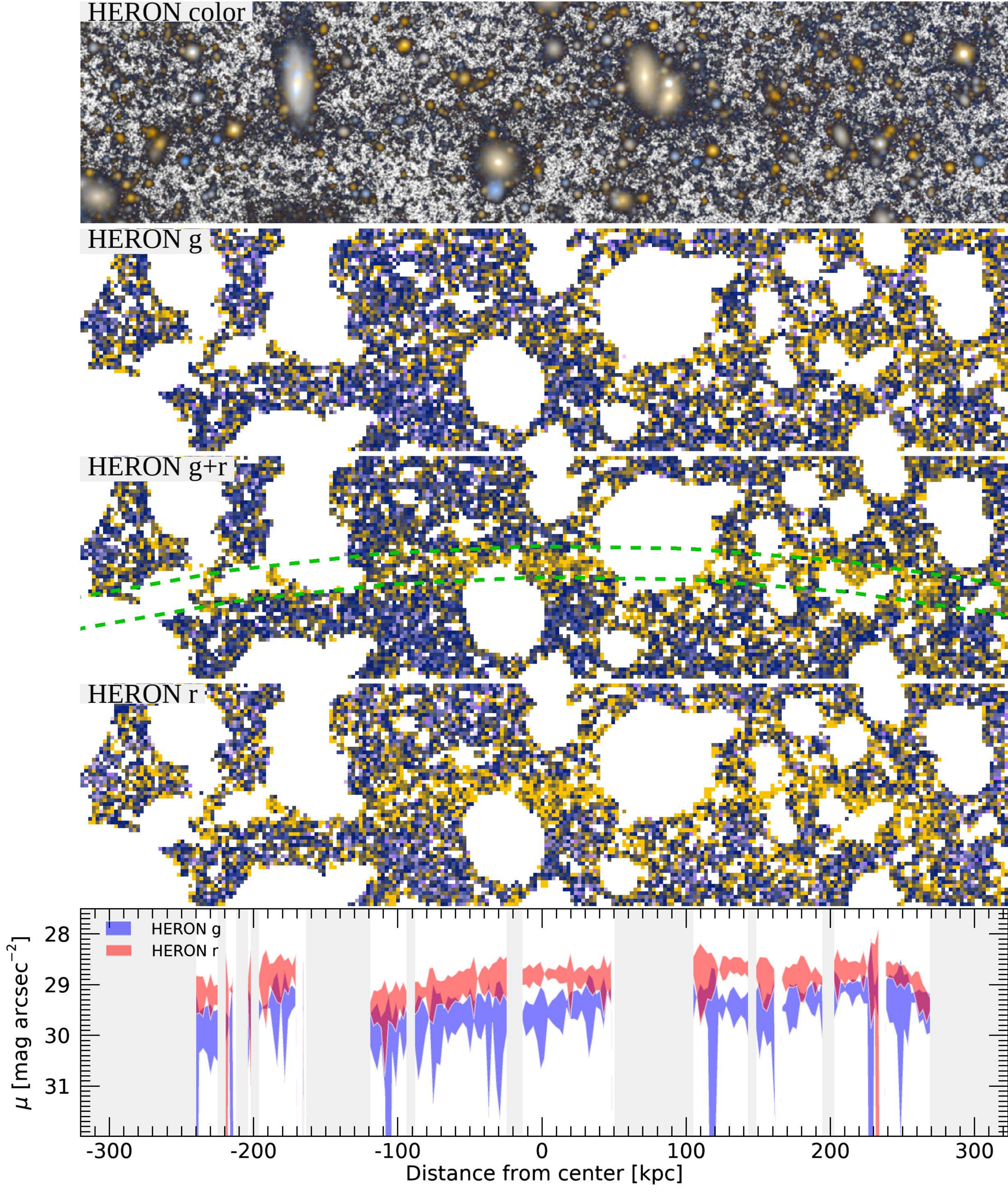}
    \caption{{Images and photometric profiles with HERON data. Top panel: Composite colour image with HERON in \textit{g} and \textit{r} bands. The high-contrast gray background is constructed with the sum \textit{g} + \textit{r} image. Middle panels: High-contrast images in the same field as the top panel in \textit{g}, \textit{g+r} and \textit{r}. Sources external to the stream have been masked and then the image has been rebinned to 5x5 pixels. The region bound by green dashed lines in the \textit{g} + \textit{r} image marks the position of the stream. Bottom panel: Surface brightness profiles of the stream along the aperture indicated in the \textit{g} + \textit{r} image. HERON \textit{g}  and \textit{r} bands are shown in blue and red colour respectively with an error of 1$\sigma$. Gray regions are spatial locations along the stream which are discarded due to the absence of signal produced by the masking from external sources.}}
    \label{fig:Photo_perfil_HERON}
\end{figure*}

We first analyzed the photometry using the HERON data. In Fig. \ref{fig:Photo_perfil_HERON}, top panel, we show a composite colour image with both \textit{g} and \textit{r} bands from HERON. This image is useful to visualize the full extent of the stream and its morphology. {The apparent size in length and width of 18.5 $\times$ 1 arcmin by visual inspection corresponds to approximately 510 $\times$ 25 kpc at the Coma distance.} In the middle panels of Fig. \ref{fig:Photo_perfil_HERON} we show the masking and subsequent binning (5$\times$5 pixels, equivalent to 5.58$\times$5.58 arcsec) of these data, allowing us to isolate the stream from external sources. We can see that the stream has a certain curvature. We estimate a radius of curvature of about 3620 arcsec, about 1 degree, equivalent to 1.7 Mpc. {The Coma virial radius is 2.4 Mpc \citep{2022NatAs.tmp..148H}.} In the middle panel of Fig. \ref{fig:Photo_perfil_HERON} we indicate with dashed lines the contours of a 45 arcsec wide annular aperture with this calculated radius of curvature. {This aperture is the width} at which the stream is detected in the HERON data, corresponding approximately to a surface brightness of 30 mag arcsec$^{-2}$ in the \textit{r} band (the shallowest of the HERON data, see Sec. \ref{sec:HERON}). This aperture is the same as the one used in Sec \ref{sec:IR} to explore the potential IR emission of the stream.

We use this aperture on the masked image to derive \textit{g} and \textit{r} band photometric profiles along the stream, which is shown in the lower panel of Fig. \ref{fig:Photo_perfil_HERON}. Because of the strong fluctuations in the profiles, {we smooth them with a Gaussian kernel} of approximately 10 binned pixel units, {in order to obtain enough signal to noise in the profile}. {We calibrate the 0 flux sky level in the region adjacent to, but sufficiently far from the stream.} We find average surface brightnesses along the stream length of approximately 29.5 mag arcsec$^{-2}$ in the \textit{g} band and 29.0 mag arcsec$^{-2}$ in the \textit{r} band. The strong fluctuations of the photometric profiles of the stream are due to the extremely low surface brightness together with the presence of masked regions, but also to possible systematic effects such as background fluctuation due to flat-fielding and sky subtraction.

\begin{figure}
\centering
        \includegraphics[width=\columnwidth]{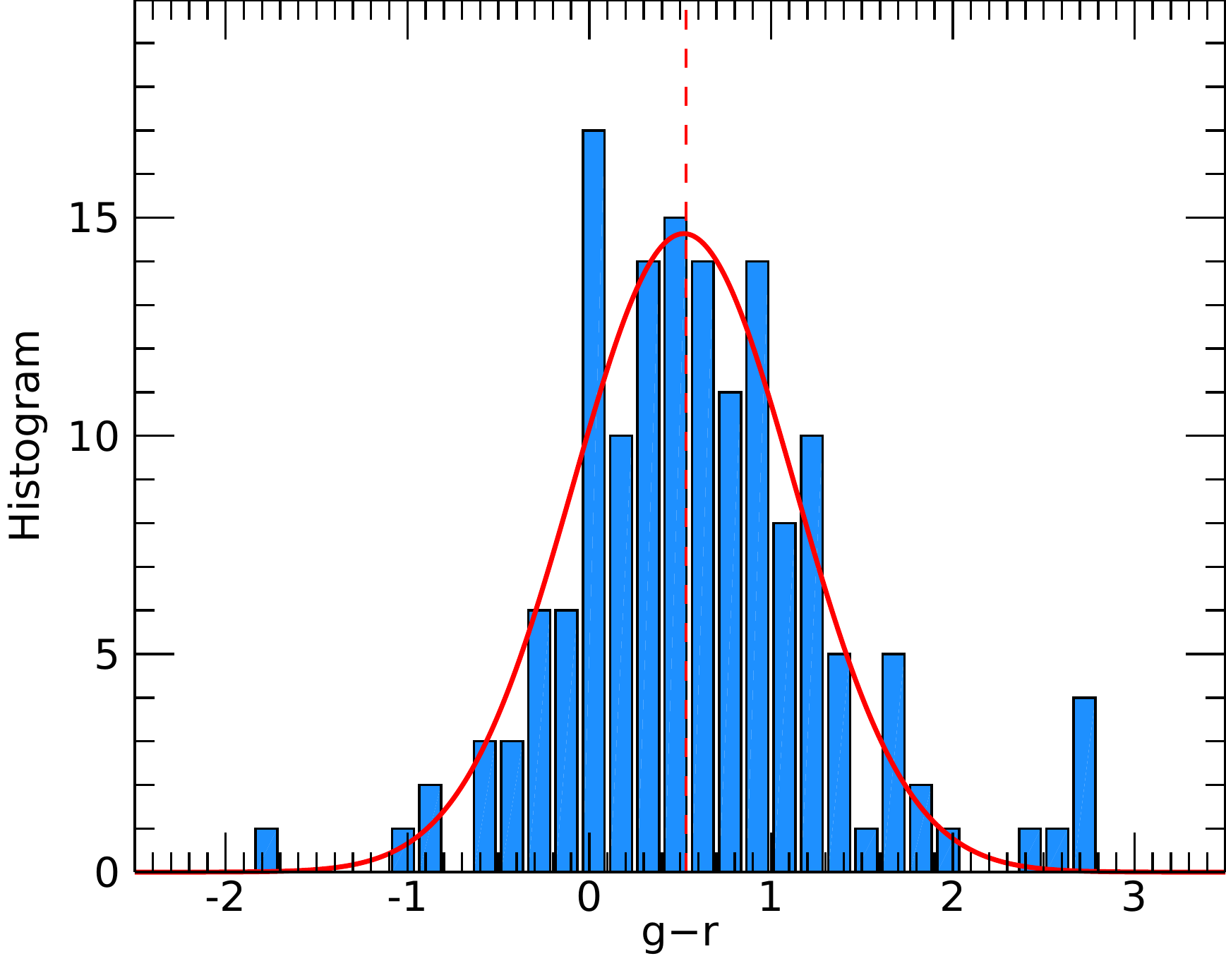}
    \caption{{Histogram of $g-r$ colour using HERON data at the selected aperture for the {Giant Coma Stream} (see Fig. \ref{fig:Photo_perfil_HERON}). The solid red line shows the best fit to a Gaussian function. The dashed red vertical line marks the mean colour calculated with a 3$\sigma$ resistant mean.}}
    \label{fig:Histogram_colors}
\end{figure}

Despite the obvious fluctuations, we observe that the difference between the \textit{g} and \textit{r} photometric profile is quite constant. In Fig. \ref{fig:Histogram_colors} we show the {pixel colour values} distribution of the stream. It faithfully follows a Gaussian function, compatible with a constant colour that would correspond to the colour of the stream. To obtain the average colour and error we calculate the mean value and error using a 3$\sigma$ robust mean, yielding $g-r$~=~0.53~$\pm$~0.05 mag. This colour indirectly indicates that a potential lensing of a high-redshift background source amplified by Coma can be ruled out, {as a high redshift source should appear much redder in $g-r$ colour, and additionally, Coma is too nearby to act effectively as a lens.} Complementary, we fit the distribution with a Gaussian function that provides an equivalent value (see Fig. \ref{fig:Histogram_colors}). {In order to obtain a pseudo colour $g-L$ with which to relate the photometric magnitudes between HERON (\textit{g} and \textit{r} bands) and WHT ($L$ band)}, we constructed an image with the average of the \textit{g} and \textit{r} bands. Using this pseudo $L$ band, we carried out a similar analysis yielding $g-L$~=~0.32~$\pm$~0.03 mag.

\begin{figure*}
\centering
        \includegraphics[width=0.95\textwidth]{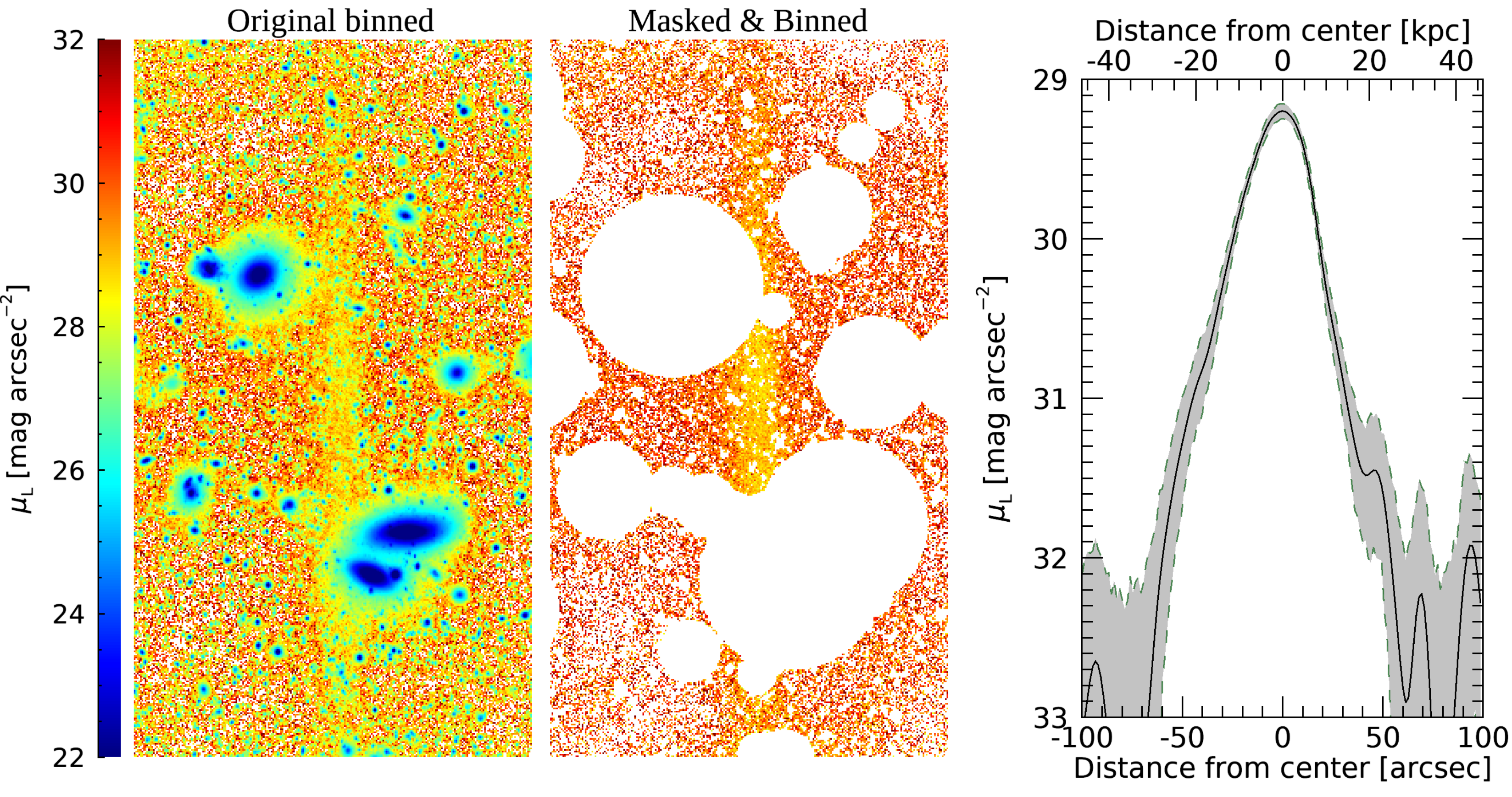}
    \caption{Surface brightness images and profiles with the WHT data in the luminance filter. Left panel: Binned image at a size of 5$\times$5 pixels (equivalent to 1.333$\times$1.333 arcsec) colour-coded in surface brightness and oriented so the stream appears vertically. Middle panel: Similar to left panel but with all external sources masked and then binned to 5x5 pix. Right panel: Photometric profile in the luminance band $L$ ({approximately equivalent} to \textit{g}+\textit{r}) in the cross section (left-right) direction. The grey error regions correspond to 1$\sigma$. The width in the transverse direction of the three panels coincide in spatial scale.}
    \label{fig:Photo_perfil}
\end{figure*}

We now focus on the analysis of the WHT data, with higher nominal depth and better resolution, but only available in the central region of the stream. We show in Fig. \ref{fig:Photo_perfil} a photometric profile along the cross-section of the stream. In the left panel we show a high contrast image binned 5x5 from the original pixel size of 0.2667 arcsec, thus at a pixel scale of 1.334 arcsec. In the central panel we show this same image with masking applied to hide external sources to the stream. {This masking is performed with a combination of \texttt{SExtractor} hiding mainly small size sources aggressively (\texttt{DETECT\_THRESH} = 0.3) and a manual masking with wide circular apertures hiding large size galaxies and their expected haloes that could appear even below the surface brightness detectable visibly in the images. After this masking, the images are binned at 1.334 arcsec/pix.}, allowing the diffuse light to emerge efficiently \citep[see][for a similar processing]{2021A&A...656A..44R}. 

Given the nearly straight morphology of the stream in this partial region, we project the combined flux in this direction, obtaining a photometric profile along the cross-section. This is done in the region of the image with a nominal depth {fainter than} 31.0 mag arcsec$^{-2}$ in the $L$ band (see Fig. \ref{fig:Mapas_depth}), avoiding the outermost areas, which due to the dithering of the observations have a lower signal to noise ratio. The whole region shown in the central panel is the one used for the photometric profile. For averaging, we use the median value along the stream direction, with errors calculated as 1$\sigma$. We introduce a tilt plane that fits the sky background in order to eliminate a small gradient present in the profile. The obtained profile is smoothed with a 3 pixel {Gaussian kernel} to eliminate small fluctuations {from the effects of masking}.

\begin{figure}
\centering
        \includegraphics[width=0.95\columnwidth]{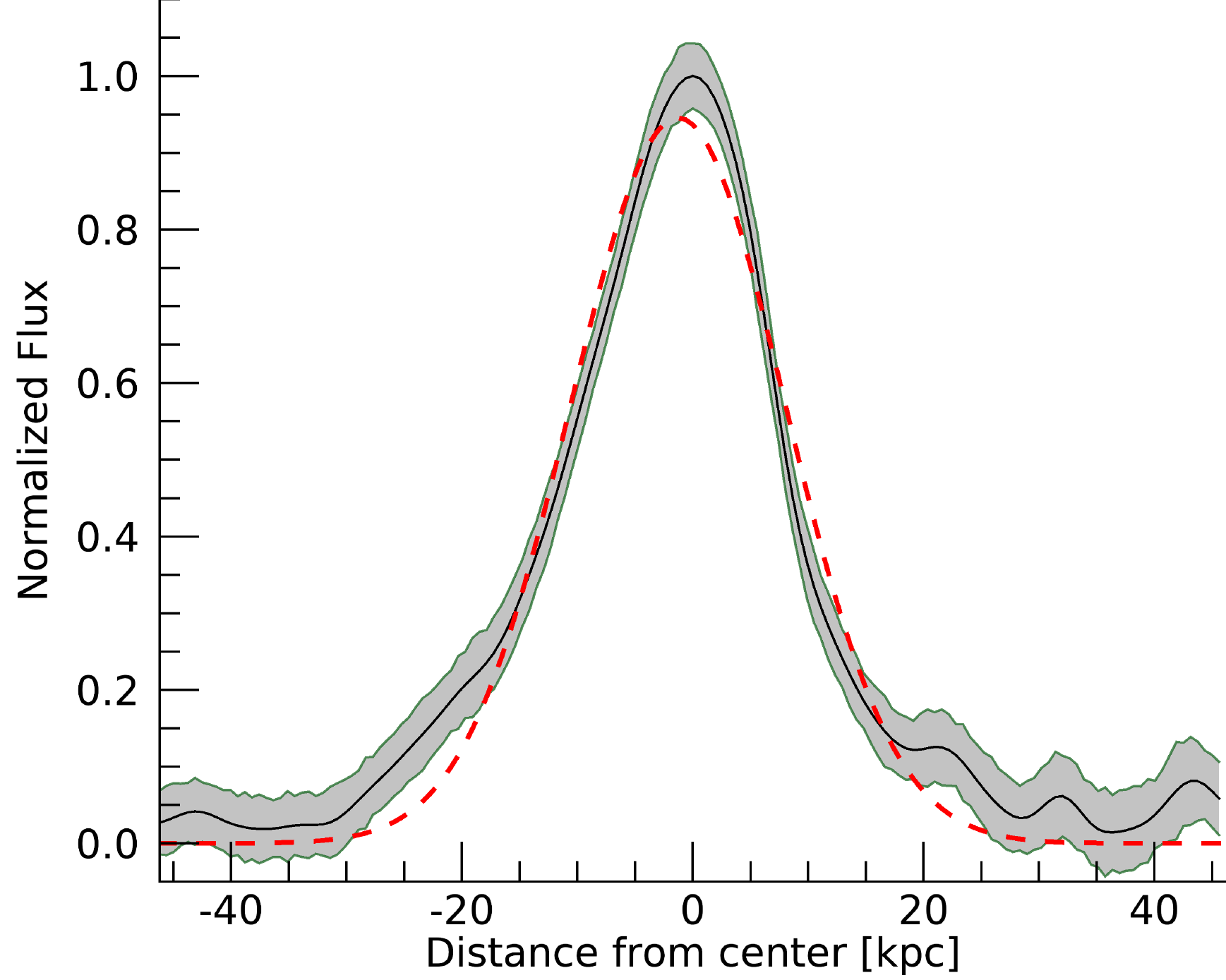}
    \caption{Photometric profile equivalent to the one in the right panel of Fig. \ref{fig:Photo_perfil} but in arbitrary flux units. The dashed red line indicates the best fit to a Gaussian function (see text). The grey error regions correspond to 1$\sigma$.}
    \label{fig:Perfil_gauss}
\end{figure}

The maximum surface brightness of the stream is approximately 29.2 mag arcsec$^{-2}$ in the $L$ band. Using the previously calculated $g-L$ colour of 0.32 mag, this would translate into a maximum surface brightness of approximately 29.5 mag arcsec$^{-2}$ in the \textit{g} band, agreeing with the values obtained from the HERON data, and giving confidence to the results. The WHT profile reaches reliably down to a surface brightness of about 32 mag arcsec$^{-2}$ where spurious fluctuations begin to show up, probably due to sky fluctuations and masking residuals. The shape of the profile {is approximately straight in its slope, therefore compatible with an exponential decay}. By analyzing this profile in flux units we can identify {an approximately} Gaussian shape (see Fig. \ref{fig:Perfil_gauss}). By modeling this flux profile with a Gaussian function, the 1$\sigma$ fitted width has a value of 20.1 arcsec or 9.3 kpc. The FWHM is 42.3 arcsec or 19.5 kpc. We notice a small asymmetry in the profile, where on the left side (towards the direction where Coma is located) the profile shows a slight excess. However, this asymmetry is not strong, potentially due to masking residuals of nearby galaxies located in that region.

Due to the numerous sources overlapping the stream, a direct flux measurement is not possible. In order to obtain an integrated magnitude we use the good fit to a Gaussian profile to model the total flux. For this we assume as average profile the one obtained in the analysis of Fig. \ref{fig:Perfil_gauss} to integrate it along the estimated length of 510 kpc (see Fig. \ref{fig:Photo_perfil_HERON}). We consider this a relatively good approximation due to the approximately constant surface brightness along the entire length of the stream. {The value produced is $L$~=~20.90~$\pm$~0.12 mag, which corresponds to $g$~=~21.22~$\pm$~0.15 mag, according to the colour $g-L$~=~0.32~$\pm$~0.03 mag calculated above. The absolute magnitude of the stream according to this model at the distance of Coma would therefore be L$_g$~=~-13.78~$\pm$~0.15 mag. Following mass to light ratio predictions by \cite[][]{2015MNRAS.452.3209R}, this {translates to a} stellar mass of $M_\star$~=~6.8~$\pm$~0.8~$\times$ 10$^{7}$~$M_\odot$. We note the potential presence of uncertainties of the integrated photometric quantities, both because of systematic errors due to modeling and possible non-detection of more structure in our data. Therefore, integrated magnitudes are to be considered as an order of magnitude estimate, and as a lower limit in mass, both the stream and its potential progenitor.}

{According to the predicted mass-metallicity relations for a galaxy of this range of stellar mass \citep{2008MNRAS.391.1117P, 2019ARA&A..57..375S}, a metallicity of approximately $Z/H$~=~-1 is expected. The colour $g-r$~=~0.53~$\pm$~0.05 mag would therefore indicate that this is an old, passive dwarf galaxy according to single stellar population models by \cite{2015MNRAS.449.1177V} \citep[see also ][]{2017MNRAS.468.4039R}.}


\subsection{Similar structures in simulations}

{To gain intuition on how such structures could arise in $\Lambda$CDM, we use the Illustris-TNG50 (TNG50 for short) simulations to insight the existence of similar streams in clusters. The TNG50 simulation has a box size $\sim 50$ Mpc on a side and a baryonic mass-per-particle $\sim 8.5 \times 10^4\; \rm M_\odot$. TNG50 is part of the Illustris-TNG suite of galaxy simulations in different volumes which all include gravity, magnetohydrodynamics and a treatment for most relevant physical {processes} involved in galaxy formation, such as star formation, metal enrichment and stellar and black hole feedback \citep{2018MNRAS.475..648P, 2019ComAC...6....2N}.}

It is difficult to estimate the frequency with which one expects this kind of thin structure like the {Giant Coma Stream} within the $\Lambda$CDM model. The formation of these thin structures will depend primarily on two circumstances: timing of the merger/infall and intrinsic properties of the progenitors. Tidal disruption that sets in too early with respect to the present time at z~=~0 at which we analyze the cluster, or progenitors that have intrinsic sizes too extended, will lead to stellar streams that are too wide to be comparable to the {Giant~Coma~Stream}. On the other hand, satellite infall that is too late or intrinsic sizes that are too compact will lead to insufficient disruption to generate stellar streams that are as long. 

While the relatively small box size in TNG50 does not allow for clusters as massive as Coma to exist in the simulated volume, the high numerical resolution needed to resolve the inner properties of the progenitors with estimated mass $M_\star$~$\sim$~$10^{8-9} \; \rm M_\odot$ makes the 50 Mpc box the most adequate to conduct our analysis. { From the TNG database, we use their friend-of-friends group information to identify spatially coherent groups \citep{1985ApJ...292..371D}, the information about haloes and subhaloes as provided by Subfind catalogs \citep{2005MNRAS.364.1105S, 2009MNRAS.399..497D} and the SubLink merger trees \citep{2015MNRAS.449...49R}.}

The most massive cluster in TNG50 (group 0) has a virial mass $M_{200} \sim 2 \times 10^ {14}\; \rm M_\odot$ and virial radius $r_{200} \sim 1200$ kpc (virial quantities are defined at the radius where the average enclosed density is $200$ times the critical density of the Universe). We note that the mass of Coma is approximately larger by a factor of 10, $M_{200}~\sim~2~\times~10^{15}\;~\rm~M_\odot$ \citep[][]{2022NatAs.tmp..148H}, so that even {analyzing the highest mass cluster in
TNG50 falls short of an appropriate comparison. Yet the use of the TNG50 simulation is justified due to the necessary mass and spatial resolution to explore the presence of such narrow streams with dwarf-mass progenitors as inferred in our calculations}.

{In order to identify similar stream structures, we select all dwarf galaxies in the stellar mass range: $M_{\star,\rm max}=10^{8-9}\; \rm M_\odot$ that interacted with our group according to the SubLink merger trees and call them progenitors. The selection is done at the time of maximum stellar mass, $M_{\star,\rm max}$. This gives us a total of $342$ progenitors for group 0}. 

We further focus our search on progenitors that have lost a significant fraction of their maximum stellar content, $f_{\star,\rm bound} < 45\%$, where the bound fraction refers to the fraction of stellar mass at $z=0$ compared to the maximum stellar mass, since we are interested in progenitors of a sizable stellar stream. {This cut is motivated to be close but smaller than half the maximum stellar mass of all satellites. In any case, the exact value of the cut is not too impactful in our results. For example, choosing all progenitors retaining 50\% of their mass would only add 2 objects to the sample, which were identified to be accreted very late and to leave remnants at much larger distances than the observed structure. We thus converged to keep our 45\% mass cut as a good proxy for our progenitors.} 

\begin{figure*}
\centering
        \includegraphics[width=1.0\textwidth]{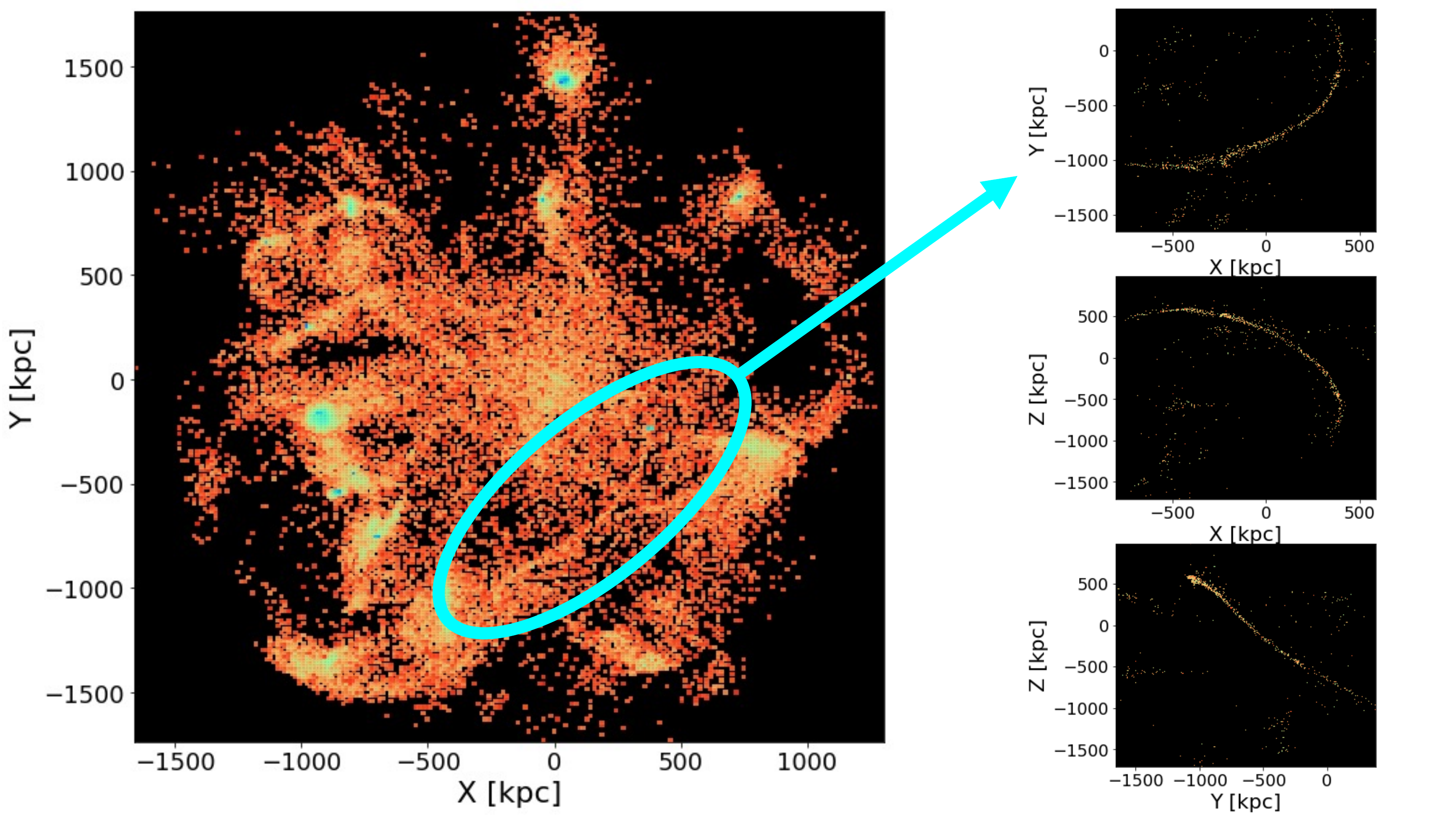}
    \caption{Projected map of stellar remnants from the identified dwarf progenitors with $M_{\star,\rm max}=10^{8-9}\; \rm M_\odot$ that fell into the most massive group simulated by the TNG50 simulation according to our selection criteria (see text). The map is colour-coded by arbitrary surface brightness units, with a resolution of 30 arcseconds. The most prominent candidate for a Coma stream analog is highlighted, and the right panels show 2D projected maps of only this remnant with a resolution of 1 arcsecond (comparable to observational resolution).}
    \label{fig:TNG50}
\end{figure*}

In addition, to identify progenitors that created structures further out in the halo of the cluster, we {require} that the median radius (r$_{50}$) of the stellar debris, { here simply defined as the material that has been tidally removed according to our substructure finder Subfind}, is larger than $800$ kpc. Our final sample comprises 19 progenitors. The main panel in Fig.~ \ref{fig:TNG50} shows an image created from the stellar debris of this sample. 

Encouragingly, within this sample, we find one particular stream with characteristics that are reminiscent of that of the {Giant~Coma~Stream}. The analogue structure is highlighted in cyan in the main panel of Fig.~\ref{fig:TNG50} and then shown in three perpendicular projections in the smaller panels on the right. Not only is this stream narrow and very extended as in the case of the {Giant~Coma~Stream}, but also under the right projection it appears as a nearly straight line (when the projection is close to the orbital plane). {We note that the centre of curvature of the stream points toward the centre of the cluster, as is the case for the {Giant~Coma~Stream}. This could be considered as further evidence that indeed the {Giant~Coma~Stream} belongs to Coma, since the approximate centre of curvature coincides with the centre of the gravitational potential about which they orbit \citep{2023ApJ...954..195N}.}

Our simulated analogue has a progenitor stellar mass $M_{\star, \rm max}= 1.1 \times 10^{8}$ and has lost $\sim 75\%$ of {its original mass} at the present day, defined at snapshot $98$ in the simulation, or redshift $z=0.009$. Interestingly, the formation history of this stream is somewhat non-canonical, as the gravitational potential of the clusters is not the only environment responsible for the stellar stream formation. The tidal disruption leading to the formation of this thin stream seems to start as part of pre-processing in a smaller group environment. Our analogue is first accreted into another group with virial mass $M_{200} = 5.48 \times 10^{11} \; \rm M_\odot$ at infall time $t_{\rm inf}=2.98$ Gyr ($z_{\rm inf}=2.2$, defined as the last time this {dwarf galaxy was centred in
its own} dark matter halo). Only at later times, by $t=6.5$ Gyr, the analog joins the friend-of-friend group of the main cluster, only crossing its virial radius by $t\sim 12.7$ Gyr, making it a relatively recent accretion. {At} the final time, the analogue has gone {through only} one pericentre around the main cluster, from which it has recently emerged and is on its way to the first apocentre. The centre of the group where the pre-processing occurred is currently more than 1000 kpc away from our analogue. 

The numerical resolution of TNG50 is not enough to study the morphology of the stream in detail, for example, in terms of width or the light profile across the stream, for which hundreds of thousands of stellar particles would be needed to faithfully trace the structure (compared to the $\sim 10,000$ available in TNG50 for this progenitor). Instead, this simulation provides a glimpse into a possible formation scenario for very narrow and long stellar streams in massive galaxy clusters, demonstrating that they do occur within the current cosmological scenario. Tailored idealized simulations would be the obvious next step to study and understand the detailed properties of the stream in terms of width, light profile and colour gradients.

\section{Discussion}

The properties of the {Giant Coma Stream} are remarkable. {Its thin and coherent morphology is reminiscent of cold stellar streams observed in the Milky Way \citep[e.g., ][]{2003AJ....126.2385O, 2006ApJ...643L..17G, 2016ApJ...820...58B} or streams in the Andromeda Galaxy \citep[see a review by ][]{2016ASSL..420..191F}. Its surface brightness peaks at a very faint level of 29.5 mag arcsec$^{-2}$ in the \textit{g} band (see Fig. \ref{fig:Photo_perfil}), comparable to the surface brightness of the Giant Stream in the Andromeda Galaxy \citep[][]{2001Natur.412...49I}.} However, the size of the {Giant Coma Stream}, with a detected length of approximately 510 kpc, is {longer} than any known {"giant" stream in the literature \citep{2001Natur.412...49I, 2009ApJ...692..955M, 2021MNRAS.506.5030M, 2023A&A...669A.103M}.} To the best of our knowledge, no stellar stream of such size has ever been detected, nor a stream of such low surface brightness by surface photometry observations.

{Previous work in Coma and other galaxy clusters revealed the presence of tidally disrupted objects \citep{1998MNRAS.293...53T, 1998Natur.396..549G, 1999AJ....117...75C, 2000MNRAS.314..324C} showing evidence for the build-up of the intracluster light by accretion of sub-structures. We discuss some particular cases. First, the plume-like object discovered by \cite{1998Natur.396..549G} in the heart of Coma is 130 kpc long and about 15-30 kpc wide. Its surface brightness is much brighter ($\mu_R$ = 25.7 mag arcsec$^{-2}$) than the {Giant~Coma~Stream}, as is its integrated luminosity (R = 15.6 $\pm$ 0.1 mag) and a much redder colour of $g$-$r$ $\approx$ 0.75 mag {\citep{1998Natur.396..549G}}. A simple calculation shows that the stellar mass of this plume-like object would range around 10$^{10}$ solar masses, indicating that the remnant is a much more massive galaxy than a dwarf, probably an elliptical galaxy, discarding a common origin. While the properties and location of both this plume-like object and the {Giant~Coma~Stream} differ, it is interesting that both share the same orientation, and that this orientation coincides in their alignment with the main filament that feeds Coma \citep[see][]{2020A&A...634A..30M}. Additionally, correlated accretion due to the orientation of filaments and large-scale structure is expected to occur within $\Lambda$CDM \citep[e.g.,][]{2014MNRAS.443.1274L}. {Another interesting object is} the arc located in the Centaurus cluster discovered by \cite{2000MNRAS.314..324C}. Its morphology is remarkably similar to the {Giant~Coma~Stream}, with a linear and featureless appearance, and a length of $\sim$ 170 kpc {\citep{2000MNRAS.314..324C}}. However, the surface brightness is considerably higher than that of the {Giant~Coma~Stream} with $\mu_R$ = 26.1 mag arcsec$^{-2}$. \cite{2000MNRAS.314..324C} argue that this object could be a tidally disrupted luminous spiral galaxy.}

{Several more recent} studies have been carried out in several galaxy clusters \citep[][]{2014ApJ...781...24G, 2018MNRAS.474..917M, 2018MNRAS.474.3009D, 2021ApJ...910...45M, 2021ApJ...922..268J, 2022ApJ...940L..51M}. The surface brightness and resolution (due to the close proximity of the Coma cluster) achieved in our work, are unprecedented, with one exception. The Burrell Schmidt Deep Virgo Survey \citep[][]{2017ApJ...834...16M} is comparable in terms of depth to our observations with the addition of a better resolution due to the proximity of the Virgo Cluster. Indeed, a large number of thin stellar streams appear in the central region of Virgo, for example a pair of streams to the NW of M87 with lengths around ~150 kpc and similarly thin morphologies and widths on the order of the tens of kpc \citep[streams A and B by][]{2005ApJ...631L..41M, 2010ApJ...720..569R}, as well as a myriad of much smaller streams or tidal features associated with different galaxies in the field of Virgo \citep[][]{2017ApJ...834...16M}. 

A fundamental characteristic that differentiates the {Giant~Coma~Stream} from the streams detected in Virgo is that it is not associated with any particular galaxy but is a free-floating structure in the external regions of Coma. Dynamically cold and extremely faint stellar streams are thought to form through strong tidal fields in low-mass accretion events \citep[][]{2005ApJ...635..931B}, and in the case of galaxy clusters through interactions of low-mass galaxies passing through the inner region of the clusters \citep[][]{2012ApJ...748...29R, 2015MNRAS.451.2703C}. These streams are fragile, and dynamical times of 1 or 2 times the crossing time are enough to destroy them \citep[][]{2009ApJ...699.1518R}. It is thus striking that the {Giant~Coma~Stream} is a free-floating stream far from the centre of the cluster, with such a coherent and fragile morphology.

We argue that the lack of detected free-floating streams in Virgo could be due to two circumstances. First, Coma is much more massive than Virgo. This implies a much higher velocity dispersion, $\sigma_v$~=~1008~km~s$^{-1}$ for Coma \citep[][]{1999ApJS..125...35S} vs. $\sigma_v$~=~638~km~s$^{-1}$ for Virgo \citep[][]{2020A&A...635A.135K}, so high-velocity galaxies are more common in Coma. Second, a free-floating stream in Virgo would be expected to lie in the periphery of the cluster, but due to the proximity of Virgo, the deep data provided by \cite{2017ApJ...834...16M} are limited to the central region of Virgo. If a similar feature exists at a projected distance of the {Giant~Coma~Stream}, 0.8 Mpc, in Virgo, that stream would be outside the footprint of these observations, and therefore undetected. This makes it potentially interesting to explore the outermost regions of the Virgo cluster, or other nearby clusters, in a search for similar free-floating streams, analogous to the {Giant~Coma~Stream}.

The presence of an analogue stream in the TNG-50 simulation may indicate that such free-floating streams with cold or thin morphology may be common in galaxy clusters. There is no guarantee that the {Giant~Coma~Stream} was formed by a complex interaction like the one depicted by our analogue. However, it is encouraging that we found at least one analogue in the formation history of a galaxy cluster. Our analysis indicates that while typical debris from this type of progenitor tends to be wider and less coherent than the {Giant~Coma~Stream}, at least there is one case where the narrowness and length of the stellar stream are comparable to it, suggesting that such structures may occur in $\Lambda$CDM. The recent accretion of this analog is compatible with the fragility of this type of structure, which, as mentioned above, is typically able to survive only 1 or 2 times the crossing time in its orbit by the cluster, making it very likely that the {Giant~Coma~Stream} is a recent accretion.

{The potential presence of such thin stellar streams of cold morphology in galaxy clusters could extend the environmental range of their study from galactic to cluster scales. The possibility of the common existence of cold stellar streams of considerably larger sizes than those found in the Local Group could make current projections of their detectability much more optimistic \citep[][]{2019ApJ...883...87P}, and future observational work, including by Euclid, the Rubin Observatory, the Nancy Grace Roman Space Telescope or ARRAKIHS, will be necessary in order to unveil similar structures in galaxy clusters and the properties of both the {Giant~Coma~Stream} and potential new discoveries.}

One of the imminent and most impactful applications of the study of cold stellar streams is related to the possibility of testing the shapes of the dark matter haloes and its subhalo distribution in their hosts, since the presence and number of subhaloes is ultimately defined by the properties of dark matter particles \citep[][]{2002MNRAS.332..915I, 2012ApJ...748...20C, 2015MNRAS.454.3542E}. Ongoing work is currently analyzing the morphology and kinematics of cold stellar streams in the Milky Way, with the objective of tracing the global gravitational potential, but also analyzing the possible occurrence of the impact of a low-mass dark matter subhalo that could perturb, both morphologically and kinematically, these streams \citep[][]{2016MNRAS.463..102E, 2019MNRAS.484.2009B, 2020ApJ...891..161I, 2021JCAP...10..043B}. 

The observational data available to us in the {Giant~Coma~Stream} do not allow a detailed analysis of this stream, due to insufficient resolution and depth. We recall that the maximum resolution of our data is approximately 1.2 arcsec FWHM, equivalent to 550 pc at the Coma distance. Additionally, radial velocity measurements of individual stars are fundamental to these analyses \citep{2022ApJ...941...19P} \citep[but see][]{2023ApJ...954..195N}. While current instrumental capabilities do not allow for such analysis, the new generation of extremely large aperture telescopes may have sufficient observational capabilities for these studies.


\section{Summary}
In this work, we report the discovery of the {Giant Coma stream}, a stellar stream with an extremely coherent and thin morphology, located at a distance of 0.8 Mpc from the centre of the Coma cluster, and which is reminiscent of the cold stellar streams detected in the Milky Way.
\begin{itemize}
    \item The properties of the {Giant Coma stream} are striking: a maximum surface brightness of $\mu_{g}$~$=$~29.5~mag~arcsec$^{-2}$ and a length of 510 kpc. This makes it the faintest stream ever detected by surface photometry. {The stellar mass is estimated to be $M_\star$~=~6.8~$\pm$~0.8~$\times$ 10$^{7}$~$M_\odot$ with a colour of $g-r$~=~0.53~$\pm$~0.05 mag, consistent with a passive dwarf galaxy as a potential progenitor for the {Giant Coma stream}}. We do not identify any potential galaxy remnant or core, and the stream structure appears featureless in our data.
    
    \item {The {Giant Coma stream} is orders of magnitude fainter in surface brightness than previous tidal features discovered in Coma and other clusters \citep{1998MNRAS.293...53T, 1998Natur.396..549G, 1999AJ....117...75C, 2000MNRAS.314..324C}. It is comparable in surface brightness to the thin streams detected in Virgo \citep{2017ApJ...834...16M}, however unlike these, the {Giant Coma stream} does not appear to be associated with any particular galaxy, appearing as a free-floating feature in the outer regions of Coma.}
    
    \item Through analysis of the Illustris-TNG50 simulation, we found a remarkably similar analogue, suggesting that such giant, {extremely faint and free-floating} thin stellar streams may {exist} in galaxy clusters according to $\Lambda$-CDM.
\end{itemize}

This work shows a glimpse of the kind of structures waiting to be discovered in the ultra-low surface brightness Universe, with promising future applications in {revealing the ongoing hierarchical assembly of galaxy clusters, as well as helping to unveil the ultimate nature of dark matter.}

\begin{acknowledgements}
{We thank the two anonymous referees for a thorough review of our work.} We thank Chris Mihos, Garreth William Martin and Claudio Dalla Vecchia for interesting discussions about the result. We acknowledge financial support from the State Research Agency (AEI-MCINN) of the Spanish Ministry of Science and Innovation under the grant "The structure and evolution of galaxies and their central regions" with reference PID2019-105602GB-I00/10.13039/501100011033, from the ACIISI, Consejer\'{i}a de Econom\'{i}a, Conocimiento y Empleo del Gobierno de Canarias and the European Regional Development Fund (ERDF) under grant with reference PROID2021010044, and from IAC project P/300724, financed by the Ministry of Science and Innovation, through the State Budget and by the Canary Islands Department of Economy, Knowledge and Employment, through the Regional Budget of the Autonomous Community. JR acknowledges funding from University of La Laguna through the Margarita Salas Program from the Spanish Ministry of Universities ref. UNI/551/2021-May 26, and under the EU Next Generation. RMR acknowledges financial support from his late father Jay Baum Rich. I.T. and G.G. acknowledge support from the ACIISI, Consejer\'{i}a de Econom\'{i}a, Conocimiento y Empleo del Gobierno de Canarias and the European Regional Development Fund (ERDF) under grant with reference PROID2021010044 and from the State Research Agency (AEI-MCINN) of the Spanish Ministry of Science and Innovation under the grant PID2019-107427GB-C32 and  IAC project P/302300, financed by the Ministry of Science and Innovation, through the State Budget and by the Canary Islands Department of Economy, Knowledge and Employment, through the Regional Budget of the Autonomous Community. The operation of the Jeanne Rich Telescope was assisted by David Gedalia and Osmin Caceres and is hosted by the Polaris Observatory Association. This work is based on service observations made with the William Herschel telescope (programme SW2021a15) operated on the island of La Palma by the Isaac Newton Group of Telescopes in the Spanish Observatorio del Roque de los Muchachos of the Instituto de Astrofísica de Canarias.
\end{acknowledgements}

%
%

\begin{appendix}

\section{Exposure times and depth in the WHT data}

\begin{figure*}
\centering
        \includegraphics[width=1.0\textwidth]{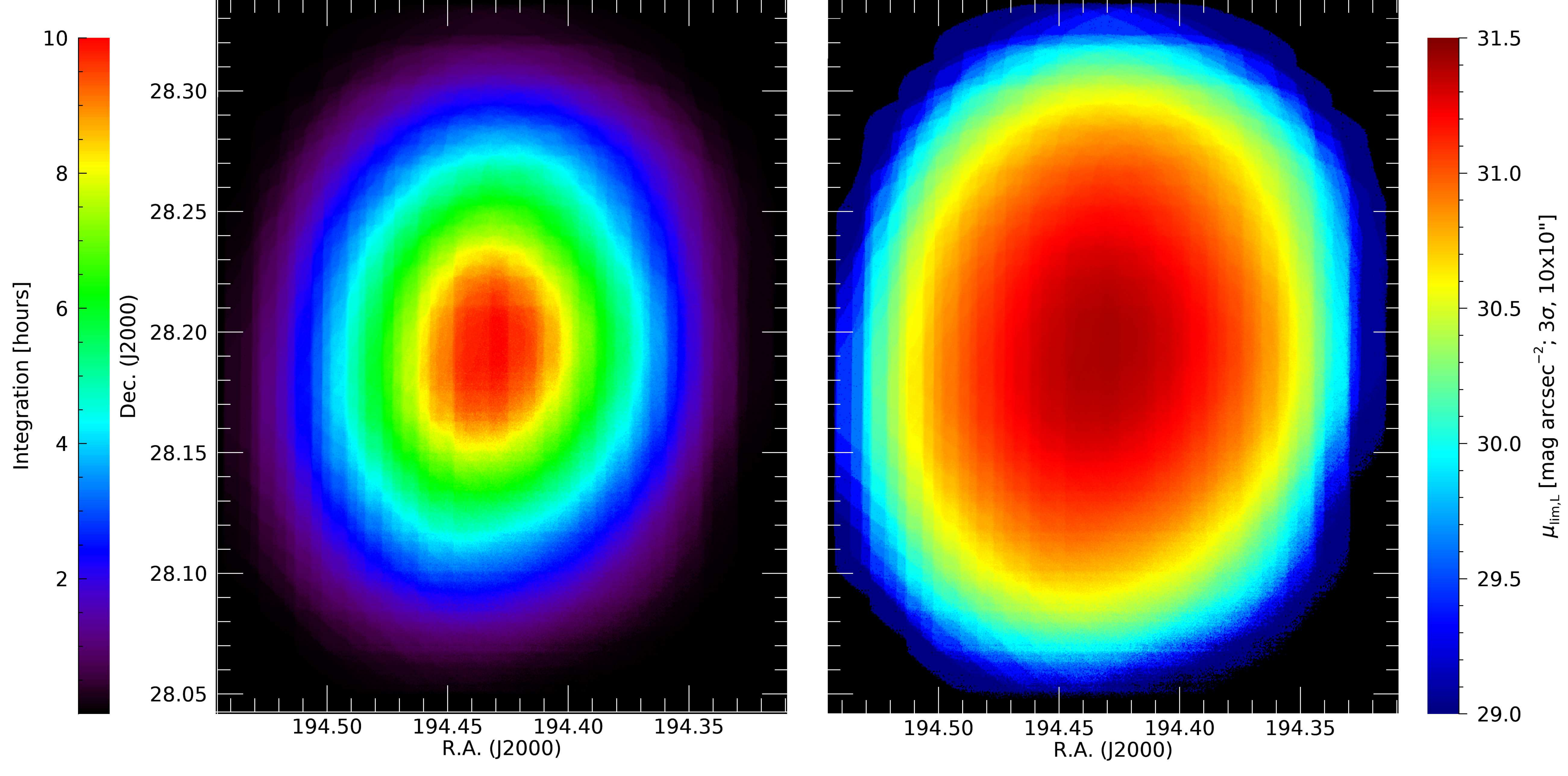}
    \caption{Maps of exposure time (left panel) and equivalent limiting surface brightness (right panel) for the WHT data in the luminance band (equivalent to \textit{g} + \textit{r} SDSS filters). Images cover a 14.15 x 17.67 arcmin rectangle. North is up, east to the left.}
    \label{fig:Mapas_depth}
\end{figure*}

\begin{figure}
\centering
        \includegraphics[width=1.0\columnwidth]{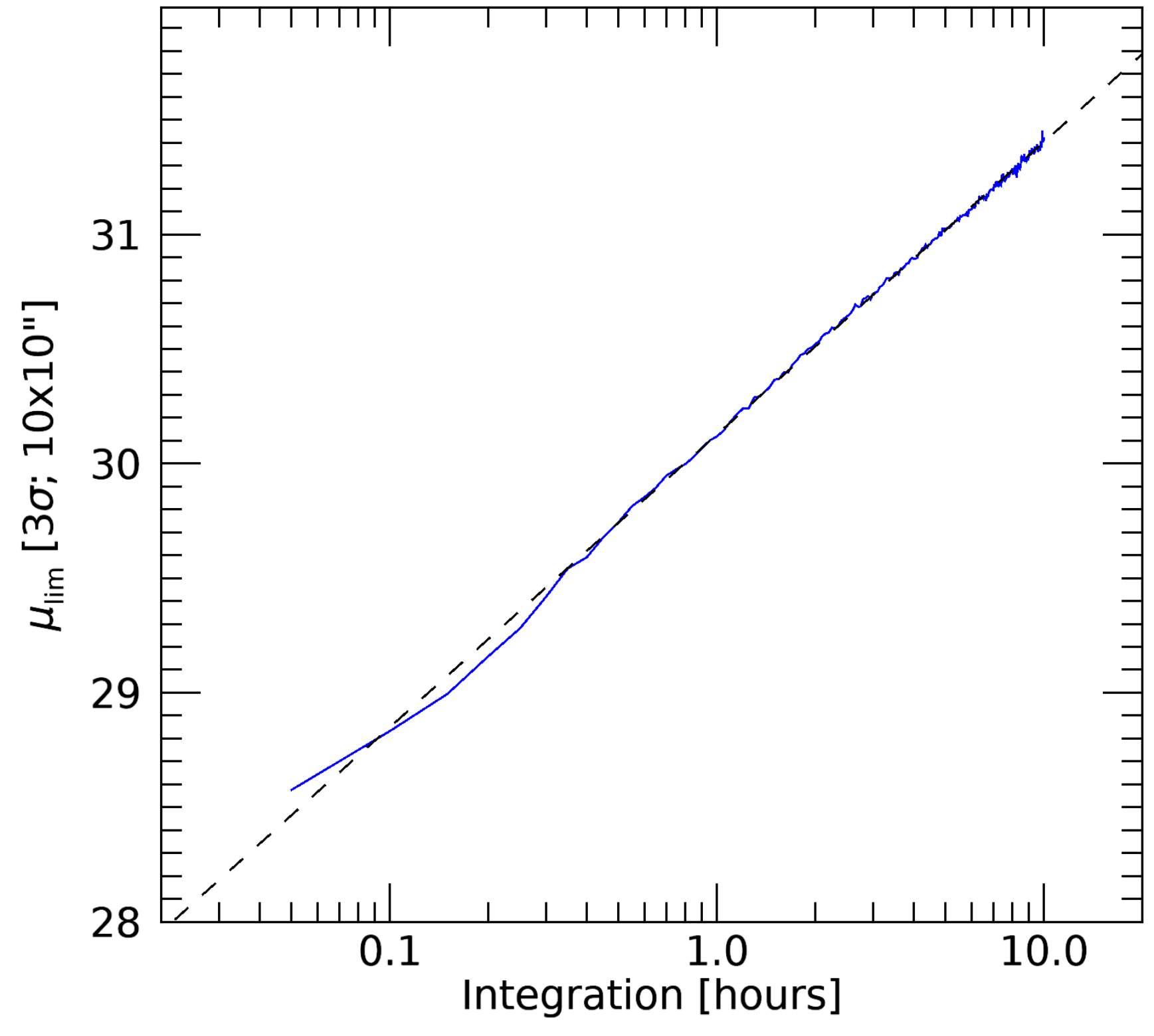}
    \caption{Linear correlation between exposure time and surface brightness limits measured as 3$\sigma$ in 10$\times$10 arcsec boxes for the WHT data. The blue line marks the correlation data obtained. The black dashed line marks the best fit of the correlation (see text). }
    \label{fig:Depth_plot}
\end{figure}

In Fig. \ref{fig:Mapas_depth}, left panel, we show the exposure time footprint for the WHT data. In order to provide a map of surface brightness limits, we perform the correlation between the standard deviation for pixels with equal exposure time and the exposure time, this correlation is shown graphically in Fig. \ref{fig:Depth_plot}. The parameters of the best fit are:

$$\mu_{lim,L} = -2.55\times log\left(\frac{1}{\sqrt{t_{exp}[h]}}\right)+30.13\;mag\;arcsec^{-2}$$

measured as 3$\sigma$ in 10$\times$10 arcsec boxes. By means of this obtained relation, we can make a surface brightness map using the exposure time values for each pixel. This map is shown in the right panel of Fig. \ref{fig:Mapas_depth}.

\end{appendix}

\end{document}